\documentclass{revtex4}
\usepackage{graphicx}
\usepackage{lettrine}

\begin{document}

\title{Topological quantum phase transitions and edge states in spin-orbital coupled Fermi gases}

\author
{Tao Zhou$^{*,1,2}$, Yi Gao$^{3}$, and Z. D. Wang$^{2}$}
 \affiliation{$^{1}$College of Science, Nanjing University of
Aeronautics and Astronautics, Nanjing, 210016, China\\
$^{2}$
Department of Physics and Center of Theoretical and
Computational Physics, The University of Hong Kong, Pokfulam Road,
Hong Kong, China\\
$^{3}$Department of Physics and Institute of Theoretical Physics,
Nanjing Normal University, Nanjing, 210023, China}

\begin{abstract}
\textbf{We study superconducting states in the presence of spin-orbital coupling and Zeeman field. It is found that a phase transition from a Fulde-Ferrell-Larkin-Ovchinnikov state to the topological superconducting state occurs upon increasing the spin-orbital coupling.
The nature of this topological phase transition and its critical property are investigated numerically.
Physical properties
of the topological superconducting phase are also explored. Moreover, the local density of states is
calculated, through which the topological feature may be tested experimentally.}
\end{abstract}

 \maketitle

\vspace{3ex}

\lettrine{T}{}he subject of topological phases in quantum systems
has been studied
intensively for the past few years due to their nontrivial properties~\cite{qi}, including
significant theoretical research interests on topological superconductors/superfluids (TSC)~\cite{roy,sch,qia,kit}.
A TSC is
characterized
by a full pairing gap in the bulk and has
topologically protected gapless states on edges of
the system. Soon after the theoretical prediction,
the TSC material was reported experimentally
through doping Cu into Be$_2$Se$_3$ (Cu$_x$Be$_2$Se$_3$)~\cite{hor,wra,andr,sasa}.

One of the most important features of TSC is
the existence of gapless edge states. Especially, the
zero energy edge states are usually related to the Majorana Femions
(MFs)~\cite{qi}, who are their own antiparticles.
MFs usually obey non-Abelian statistics and have a potential application in quantum computation,
so that the realization of MFs is of significant interest.

As is known, a typical TSC is the $p+ip$ pairing system~\cite{qi}.
Besides the $p$-wave pairing, the TSC could also be realized in the system with an $s$-wave pairing plus the spin-orbital coupling and spin polarization.
Actually, it has been shown that the $s$-wave pairing system with a strong spin-orbital coupling can be equivalent
 to a $p+ip$ system~\cite{lfu,sato,zhcw,jliu,rwei}.
 Thus, apart from the intrinsic TSC material such as Cu$_x$Be$_2$Se$_3$, there may also exist some
  artificially-made TSCs.
 A kind of  candidate systems are ultracold atoms, in which
 both the $s$-wave superfluidity and spin-orbital coupling have been realized experimentally~\cite{bou,chin,lin,wang,cheu}. Meanwhile,
 it was also proposed theoretically that the topological phase and MFs can be realized
  in a strong spin-orbital coupled material,
 through coupling to an $s$-wave superconductor and an effective Zeeman field~\cite{lutc,oreg}. Recently, such kind of proposal has been realized experimentally and the signatures of MF excitations
 have been
 reported by several groups~\cite{wil,rok,den,das,mou}.

Another quite arresting feature
 in superconducting
systems seems to be the possible appearance of Fulde-Ferrell-Larkin-Ovchinnikov (FFLO) states. This kind of states
were predicted theoretically in 1960s~\cite{ful,lar} for some superconductors
subject to a strong magnetic field, namely, the Cooper pairs have a finite momentum and thus the order parameter varies periodically
in real space.  However,  this long-sought inhomogeneous superconducting state has not been observed experimentally for quite long time, possibly due to the impurity effect~\cite{asl} or the orbital effect induced by the field~\cite{grue}.

For low dimensional superconducting systems, the orbital effect is negligibly weak when the magnetic field is parallel to the superconducting layers. Thus they can be promising candidates
for realizing the FFLO states.
  In the past decade,
indications for possible FFLO states have been reported in several families of materials~\cite{rado,bia,kum,tan,uji,bal,shi,man}. Meanwhile, a signature of the FFLO state was also observed in
one-dimensional partially spin polarized cold atoms~\cite{liao}. The properties of FFLO states have attracted renewed interest due to the experimental breakthrough.

As mentioned above, the TSC may be generated in  $s$-wave superconductors with the additional spin orbital coupling and Zeeman field. While the conventional FFLO state is also expected to be generated by applying a Zeeman field
 on $s$-wave superconductors, the difference between them is whether  the spin-orbital coupling exists or not. Thus it is quite intriguing to study the effect of the spin-orbital coupling on the FFLO state~\cite{han-wang}. This issue has attracted considerable interest very recently~\cite{kseo,chqu,cqu,yxu,zha,fwu}.
 The spin orbital effect would break the FFLO modulation and the topological phase shows up for a moderate-strength spin orbital coupling.
 Generally, there may exist a phase transition from the FFLO state to the TSC state. The nature of such transition and physical properties near the critical point are
 of fundamental interest. Moreover, they should be quite important for exploring potential applications of topological phases.
 In this paper, motivated by the above consideration,
we study theoretically the interplay between the spin orbital effect and the exchange field. Based on a lattice model, the nature of the phase transition from the FFLO to TSC is explored numerically. Our results show that there exist two regions of topological states with the chemical potential being $\mu$ near $-4$ or $0$. Interestingly, near the region of $\mu=0$ , there are two effective energy bands crossing the Fermi energy. This is essentially different from the case of the continuous model~\cite{kseo}. The existence of the two bands is important to realize the FFLO state.
Thus we propose that the phase transition from an FFLO state to a topological state may be easier to occur in the lattice model.
The properties of topological phase and the MF excitations are also investigated in detail. We also study the local density of states (LDOS) to compare with the experiments.

\vspace{3ex}

\noindent
\textbf{Results}

\noindent
We start from a model Hamiltonian that includes the spin-orbital coupling, the Zeeman field coupling, and the $s$-wave pairing term, which is given by
\begin{eqnarray}
H=-&\sum_{\bf i}[\psi^{\dagger}_{\bf i}({{\sigma_0}-i
\lambda{\sigma_1}})\psi_{{\bf i}+\hat{x}}+\psi^{\dagger}_{\bf i}({{\sigma_0}-i\lambda{\sigma_2}})\psi_{{\bf i}+\hat{y}}+h.c.]\nonumber\\
&-\sum_{\bf i} \psi^{\dagger}_{\bf i}(\mu\sigma_0+h\sigma_3)\psi_{\bf i}\nonumber\\
&+\sum_{\bf i}(\Delta_{\bf i} \psi^\dagger_{\bf i} i\sigma_2 \psi^\dagger_{\bf i}+h.c.),
\end{eqnarray}
with $\psi_{\bf i}=(\psi_{{\bf i}\uparrow},\psi_{{\bf
i}\downarrow})^\mathrm{T}$, where $\sigma_n$ are the identity (n=0)
and Pauli matrix $(n=1,2,3)$, respectively. $\lambda$ is the
spin-orbital coupling strength and $h$ represents an effective Zeeman
field. Here we study different phases and the phase transition at zero temperature.
If not specifically indicated,
the parameters are chosen as $h=0.6$ and $\mu=0$; $\Delta_i$ is obtained self-consistently with the pairing potential $V=2.4$ (See methods).

We first illustrate numerically the topological feature of the present model.
In the topological state, if we consider the
edge effect to occur along the $x$-direction and the order
parameter to be uniform along the $y$-direction,
the two-dimensional Hamiltonian can be reduced to a quasi-one dimensional model [See methods, Eqs.(5-7)].
The self-consistent results with the lattice size $400\times 400$ are shown in Fig.1. Here the periodic and open boundary conditions along $y$- and $x$- directions are used, respectively.

\begin{figure}
\centering
  \includegraphics[width=3.3in]{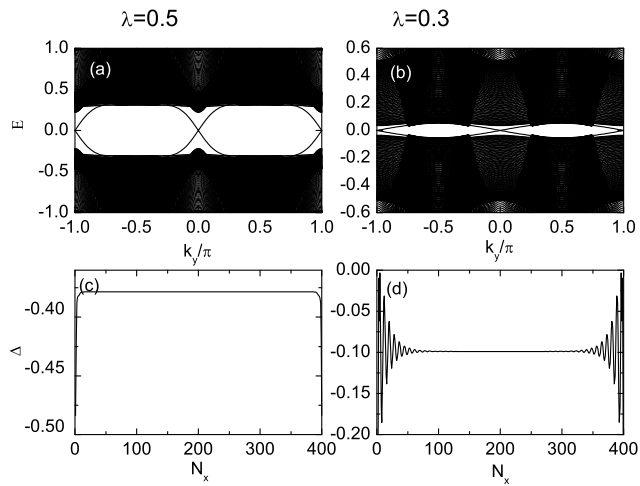}
\caption{The topological feature for a two dimensional system. (a) and (b) are the numerical results for the energy spectra with different spin orbital coupling strengths, respectively. The open and periodic boundary condition along the $x$- and $y$- directions are considered.
(c) and (d) are the corresponding order parameters. The other parameters are $\mu=0$ and $h=0.6$. }
\end{figure}

The energy bands
are plotted in Figs.1(a) and 1(b). As is shown, for both
spin-orbital coupling strengths we considered, there exists an energy
gap and one gapless state. We have checked numerically that the
spatially distribution of the gapless states are at the system edge. The existence of the edge states indicates the topological feature of our present model.
The gap magnitude depends on the
spin-orbital coupling strength; it decreases significantly as the spin orbital coupling decreases.

The order parameters with different spin-orbital
coupling strengths are plotted in Figs.1(c) and 1(d). As is seen, they
are uniform in the bulk and exhibit a boundary effect
near the edge.
For the stronger spin orbital coupling strength $\lambda=0.5$,
an obvious
boundary effect  occurs only within the ten lattice constants away from the edge [Fig.1(c)].
While when $\lambda$ decreases, the size of more lattice constants is affected by the boundary. For $\lambda=0.3$, as is seen in Fig.1(d), the order parameter
oscillates significantly, with the magnitude of the oscillation decreasing when away from the boundary. At more than
one hundred lattice constants from the boundary, the order parameter recovers to the bulk one. Previously, it was also reported that the wave function may oscillate
near the vortex core and system edge~\cite{lmao}. As proposed in Ref.~\cite{lmao}, it might make it difficult to realize a single qubit gate for universal topological quantum computation using the tunneling between two vortices.
On the other hand, such oscillation is of fundamental interest. As the spin-orbital coupling decreases further, we expect that the oscillated lattice should increase further and expand to the whole system at certain critical spin orbital coupling strength. Then the system will transit to the FFLO state for weaker spin orbital coupling.
The above expectation will be verified by the numerical results in the quasi-one-dimensional lattice, which will be presented in the following.

\begin{figure}
\centering
  \includegraphics[width=3in]{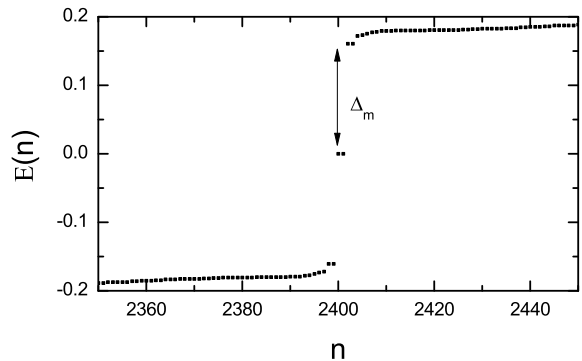}
\caption{The low energy spectrum for the quasi-one-dimensional system with the open boundary condition along the $x$-direction. $n$ is the index of eigenstates. Two zero energy eigenvalues can be seen clearly, protected by a minigap $\Delta_m$. The parameters are $\lambda=0.5$, $\mu=0$, and $h=0.6$.}
\end{figure}

To conduct a more systematic and accurate numerical study on the nature of the phase transition,
we perform numerical calculations on the three-leg ladder system (with the size $400 \times
3$) in the following. The low energy eigenvalues of the Hamiltonian are plotted in Fig.2 with $\lambda=0.5$. Here we use a
fully self-consistent calculation [Eqs.(3) and (4) in methods section], with the open
boundary condition along the $x$-direction. As is seen in Fig.2, there exist two degenerate zero-energy eigenvalues protected by a minigap about 0.16.
In the BdG framework, the eigenvalues $\pm E$ usually come from one physical particle. In the case of the zero energy mode, one physical particle  corresponds to one pair Majorana Fermions. The two MFs are spatially separated and protected by the minigap, thus they cannot be removed by local weak perturbations.
The existence of the protected zero-energy state also confirms the topological feature of the system with the present parameter.

\begin{figure}
\centering
  \includegraphics[width=3.4in]{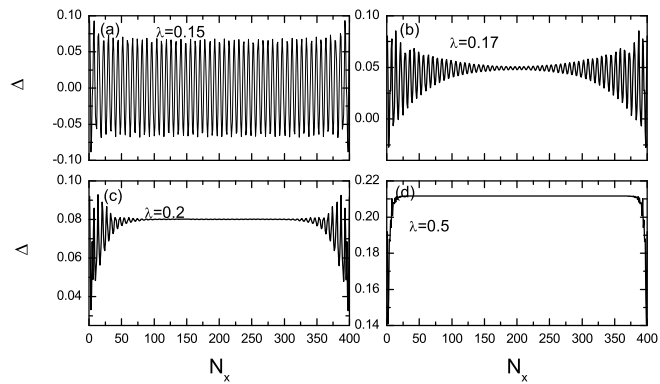}
\caption{The numerical results of the order parameter $\Delta$ in different phases for the $400\times 3$ system.
(a) The order parameter in the FFLO state. (b) The order parameter at the critical point. (c-d) The order parameter in the topological state.
The parameters are $h=0.6$ and $\mu=0$.}
\end{figure}

We elucidate numerically the nature of the phase transition driven by the spin-orbital coupling. The order parameters with different spin-orbital strengths
are plotted in Fig.3. As is seen from Fig.3(a),
 the order parameter oscillates periodically in the whole space for the small spin-orbital strength $(\lambda=0.15)$. Such oscillation is different from the above one shown in the topological state [Fig.1(d)].
 Firstly,
 the order parameter changes the sign within one periodical lattice, thus the phase of the order parameter changes and also varies periodically. Secondly, the magnitude of the oscillation does not change in the bulk. So the system is in the FFLO state for this weak spin-orbital coupling.
 As $\lambda$ increases to $\lambda=0.17$, the result of order parameter changes qualitatively. As is seen, there exists an obvious boundary effect, namely, the order parameter oscillates strongly near the boundary while it oscillates very weakly in the bulk. The phase of the order parameter does not change when away from the boundary. On the other hand, the order parameter still oscillates in the whole space~\cite{note}. Thus this coupling strength $\lambda=0.17$ should be the critical coupling strength between the FFLO state and the TSC phase. As $\lambda$ increases further [Fig.3(c)], the oscillation region decreases and the order parameter is uniform in the bulk. For a larger spin-orbital strength, the oscillation behavior disappears and the order parameter recovers to the bulk value within several lattice-constants away from the boundary. The above results of the TSC state are qualitatively the same as those of the two-dimensional lattice shown in Fig.1.

\begin{figure}
\centering
  \includegraphics[width=3.4in]{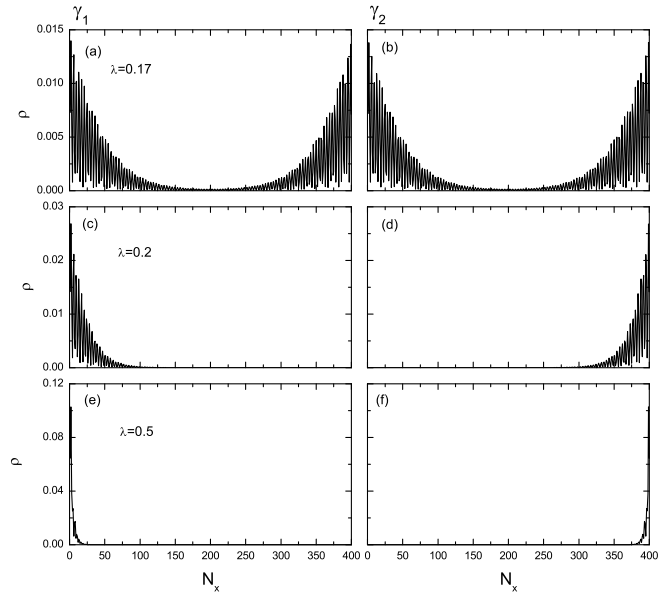}
\caption{The spatial distributions of the two MF states with different spin-orbital coupling strengths.
(a) The MF states at the critical point where the FFLO modulation disappears. (b) The oscillated MF states in the topological state.
(c) The spatial distribution of the MF states when entering deeply into the topological state. The parameters are $h=0.6$ and $\mu=0$.
}
\end{figure}

Now let us study the properties of the TSC phase. As shown in Fig.1,  these exists a gapless edge state in this phase for the two-dimensional systems.
For the present model, there should be two MFs associated with the zero energy state, expressed as $\gamma_1$ and $\gamma_2$.
With a zero-energy fermion $C^{\dagger}=\sum_{i\sigma}(u_{i\sigma}\psi^{\dagger}_{i\sigma}+v_{i\sigma}\psi_{i\sigma})$, $\gamma_1$ and $\gamma_2$ are expressed as:
$\gamma_{1}=(C+C^{\dagger})/\sqrt{2}$ and $\gamma_{2}=i(C^{\dagger}-C)/\sqrt{2}$. Here the zero-energy fermion $C^{\dagger}$ can be obtained from the BdG equation,
 thus the MF states can be studied numerically. The numerical results of the spatial distribution of the two MF states are presented in Fig.4. At the critical point, as is seen in Figs.4(a) and 4(b), the distributions of the two MFs are the same so that they shall annihilate to an ordinary fermion. The distribution curve oscillates in the whole lattices. And the oscillation is stronger near the boundary. This is similar as the case of the order parameter.
As the spin orbital strength increases to $\lambda=0.2$, the two MFs are completely separated and locate near the two boundaries, respectively. The distribution curve still oscillates with the depth about 100 lattice size. As the spin-orbital strength increases further, the oscillation disappears and the two MFs are located near the boundaries. These results may be verified by experiments and may be useful when exploring the application of MFs in the topological quantum computation.

Finally, let us look into the LDOS to disclose
the existence and distribution of the zero
mode. The LDOS spectra as a function of the energy at the boundary for different spin orbital coupling are plotted in Fig.5(a). As is shown,
at the critical point ($\lambda=0.17$), one can see a weak zero energy peak. As the spin-orbital strength increases, the intensity of the zero energy peak increases. For the strong strength ($\lambda=0.5$), the spectra weight of the zero energy peak is quite strong. Actually, the zero bias peak has also been observed by very recent experiment and it was believed to be signature of the MFs~\cite{mou}.
The LDOS spectra as a function of the real space position are presented in Figs.5(b-d). Here the LDOS spectra may be qualitatively consistent
with the distributions of the two MF states.
 This is understandable because the contributions from non-zero energy spectra are very small due the existence of the minigap. The LDOS spectra also oscillates near the boundary for $\lambda=0.17$ and $\lambda=0.2$, which might be verified by later experiments. The numerical results for LDOS may establish a useful link for theoretical
calculations and experimental observations.

\begin{figure}
\centering
  \includegraphics[width=3.4in]{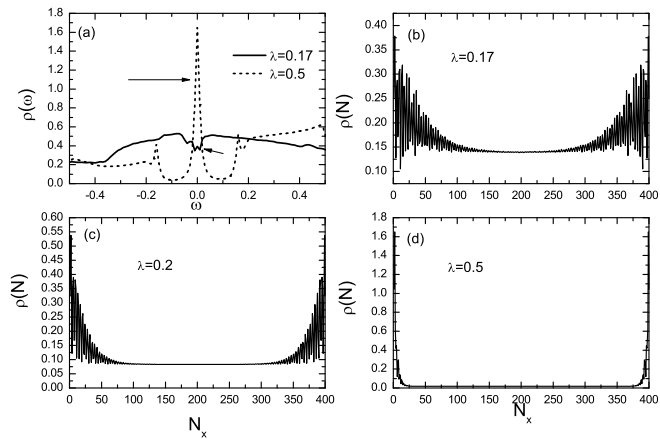}
\caption{ The numerical results of the LDOS. (a) The LDOS as a function of the energy. The zero energy states are indicated by the arrows. (b-d) The LDOS as a function of the spatial position with different spin-orbital coupling strengths. The parameters are $h=0.4$ and $\mu=0$. }
\end{figure}

\vspace{3ex}

\noindent
\textbf{Discussion}

\noindent
We can explain qualitatively the origin of the phase transition and the oscillation of the order parameter in the topological phase.
In the momentum space, the two dimensional normal state Hamiltonian $(\Delta_i=0)$ can be rewritten as $H=\sum_{\bf k}(\psi^{\dagger}_{\bf k\uparrow},\psi^{\dagger}_{\bf k\downarrow})H({\bf k})(\psi_{\bf k\uparrow},\psi_{\bf k\downarrow})^{\mathrm{T}}$, with
\begin{equation}
H({\bf k})=
\left(
\begin{array}{cc}
\varepsilon_{\bf k}-h & -2\lambda(\sin k_x-i\sin k_y) \\
-2\lambda(\sin k_x+i\sin k_y) &\varepsilon_{\bf k}+h
\end{array}
\right),
\end{equation}
where $\varepsilon_{\bf k}=-2(\cos k_x+\cos k_y)-\mu$. Without the spin orbital coupling ($\lambda=0$), the normal state energy bands of the spin up and spin down electrons are $\varepsilon_{\bf k}\pm h$, respectively. As a result, the Fermi surfaces of the spin up and spin down electrons are separate. For the case of $s$-wave symmetry, only inter-band pairing is available. Since the Fermi momentum of spin down electron is less than that of spin up electron, there exists a net momentum for the Cooper pair.
Thus the order parameter oscillates periodically in the real space.

In the presence of the spin orbital coupling, the renormalized normal state energy bands are expressed as
\begin{equation}
E_\pm=\varepsilon_{\bf k}\pm\sqrt{h^2+4\lambda^2(\sin^2 k_x+\sin^2 k_y)}.
\end{equation}
The quasiparticle operators can also be obtained
through diagalizing the Hamiltonian [Eq.(2)]. Both are expressed as the superposition of the spin up and spin down electrons. As a result, the intra-band pairing becomes possible, with the weight being determined by $\lambda$.
The inter-band pairing would generate the FFLO-type modulation. The intra-band one would generate zero-momentum Cooper pair. And here
the Cooper pairs are made up from two spinless quasiparticles, so that the case of intra-band pairing is equivalent to two-band $p$-wave superconductor with opposite chirality.

Let us first brief an idea about the topological phase based on the continuous-type model~\cite{jdsau,jsau,alic}, with the band dispersion expressed as $k^2/2m-\mu\pm \sqrt{h^2\pm \alpha \mid {\bf k}\mid^2}$. A basic point to obtain a topological phase is that the Zeeman field splits the two spin-orbital bands with a gap $2h$.  One may choose the chemical potential to place the Fermi level inside the gap. As a result, the electrons only occupy the lower band, while the upper unoccupied band almost plays no role. Based on the above point, the parameter region for the appearance of the topological phase has been obtained. One can conclude from the band dispersion that the Fermi level only crosses the lower band when the chemical potential is less than the exchange field ($\mid \mu\mid<h$). Thus the system is equivalent to a one-band $p+ip$ pairing system.
With a pairing gap $\Delta$ and an effective exchange field $h$,  the parameter region for the topological phase is $h>\sqrt{\Delta^2+\mu^2}$~\cite{jdsau,jsau,alic}.

\begin{figure}
\centering
  \includegraphics[width=5.4in]{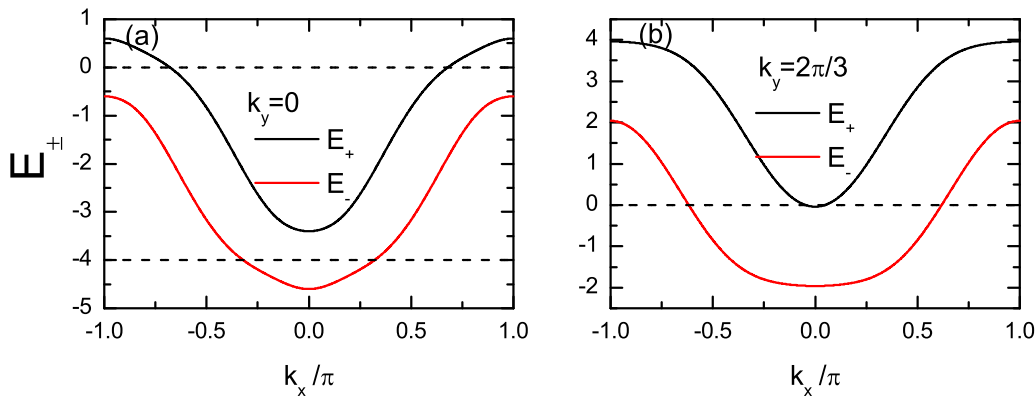}
\caption{The quasiparticle band dispersion [Eq.(3)] along the cuts $k_y=0$ and $k_y=2\pi/3$, respectively. The parameters are: $\lambda=0.5$, $h=0.6$, $\mu=0$.}
\end{figure}

For the lattice-type model, the energy bands are expressed by Eq.(3). The topological behavior should be determined by the four points $(0,0)$, $(\pi,0)$, $(0, \pi)$, and $(\pi,\pi)$. For the case of quasi-one dimensional system with finite odd lattice size along $y$ direction, the topological behavior is determined by the $\Gamma=(0,0)$ and $X=(\pi,0)$ points.
We plot in Fig.6 the band dispersions for $E_{\pm}$ along the $k_y=0$ and $k_y=2\pi/3$ cuts to discuss the nature of the topological phase in our system.
Here the $\Gamma$ point is the bottom of the energy bands with the energies $-4\pm h$. If one sets the Fermi level between them, then only the lower band crosses the level. As a result, this system is equivalent to a one-band $p+ip$ system, similar to the continuous model as discussed above. The topological state near $\mu=-4$ has been studied previously~\cite{sato} and is not concerned with in the present work.
On the other hand, the energy reaches the maximum at $X$ point along $k_y=0$ cut. If we set the chemical potential to satisfy $\mid \mu\mid<h$, then the Fermi level is also inside the gap between the lower band and upper band. With a particle-hole transformation, the upper band is occupied with holes, while the lower band is unoccupied. This leads to the topological feature of the system. However, here the system is indeed an effective two-band system. Actually, the lower band also crosses the Fermi level along the other direction of the Brillouin zone, e.g., along the $k_y=2\pi/3$ direction, as seen in Fig.6(b). This would not affect the topological feature of the system. The existence of two sheets of Fermi surface for the case $\mu=0$ is essential for the appearance of the FFLO state when the spin-orbital coupling strength decreases or disappears.

Then one can understand
the nature of the phase transition. For a two-band system with the bands expressed by Eq.(3),
the FFLO state comes from the interband pairing. The topological state come from the intra-band pairing. The origin of the phase transition is due to the competing of these two kinds of pairing. As the spin-orbital coupling increases, the intra-band pairing dominates over the inter-band one, so that the long range FFLO order is broken and the phase transition occurs. However, even in the topological phase, the local FFLO modulation may still survive as the spin orbital coupling is not strong enough. This will make the order parameter oscillate near the edge, as shown in Fig.3.
\begin{figure}
\centering
  \includegraphics[width=5.4in]{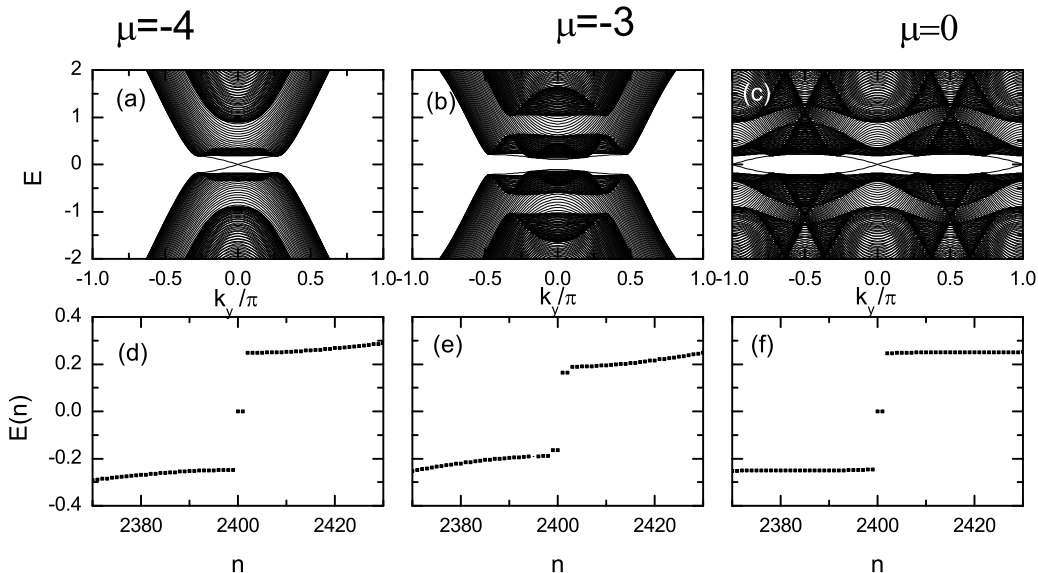}
\caption{The topological state for the chemical potential near $-4$ and $0$. (a-c): The energy spectra for the $100\times100$ two dimensional lattice with different chemical potentials. (d-f): The quasiparticle energies for the $400\times3$ lattice. The open and periodic boundaries are considered along the $x$ and $y$ directions, respectively. The parameters are set as $h=0.6$, $\Delta \equiv 0.3$, and $\lambda=0.5$.}
\end{figure}

The appearance of the topological state for the chemical potential near $-4$ and $0$ can be checked numerically. Here we set the pairing gap $\Delta$ and the exchange field $h$ to be $\Delta=0.3$ and $h=0.6$. We consider the open (periodic) boundary conditions along the $x$ ($y$) direction. The energy bands for the $100\times 100$ two dimensional lattice and $400\times 3$ quasi-one-dimensional lattice are presented in Fig.7. As is seen in Figs.7(a-c), for the chemical potential $\mu=-4$, the energy bands are gapped in the bulk while the gapless edge state exists at the $k_y=0$ point, indicating that the system is in the topological state. The gapless state disappears for $\mu=-3$, thus the system is in the topological trivial state. For the case of $\mu=0$, the gapless states appear at $k_y=0$ and $k_y=\pi$. These two gapless mode are contributed by the $(\pi,0)$ and $(0,\pi)$ points. So there are two topological regions.
We have also checked numerically that the system is in the tropological state if the parameters satisfy $h>\sqrt{(\mu+4)^2+\Delta^2}$, or $h>\sqrt{\mu^2+\Delta^2}$.
For the quasi-one dimensional lattice, as seen in Figs.7(d-f), the zero-mode exists for $\mu=-4$ and $\mu=0$. Both are protected by a minigap. We have also checked numerically that the parameter regimes for the topological state are the same as the case of the two-dimensional lattice. And here we focus our study on the region $h>\sqrt{\mu^2+\Delta^2}$.
 Our main results shown in Figs.(1-5) are robust for this parameter region, which has been checked numerically (not presented here).

Recently, the interplay between the FFLO state and topological state in the presence of the $c$-axis Zeeman field has also been studied~\cite{kseo,chqu}.
We now compare our results with theirs. A primary difference has been revealed in our above discussion, i.e.,
there exist two bands crossing the Fermi level in the topological state. As a result, as the spin-orbital coupling strength decreases, the topological state can smoothly connect to the
FFLO state and a second order phase transition occurs.
For the model used in Refs.~\cite{kseo,chqu}, the realization of the topological state requires that the Fermi level crosses only lower band.
While the FFLO state requires two separated Fermi surfaces. In order to obtain a direct transition from the FFLO state to the topological state,
  a very strong pairing potential is usually required to overcome the gap between the Fermi level and the upper band. Thus the phase diagram should be quite different from ours. And the large pairing gap would also prevent the topological state to appear. Generally, there may exist a phase transition from the FFLO state to the topological trivial superconducting state. The direct transition from the FFLO to topological state may only occur for very large pairing potential and very large exchange field.
Besides this important difference, there are some other technique differences between our work and others. Namely,  it was argued in Ref.~\cite{kseo} that the study of one-dimensional system is problematic according to the Mermin-Wagner theorem~\cite{kyang,fid,jayd}. Thus the coupling chains are considered.
The FFLO modulation occurs only along the chain direction and the order parameter is assumed to be uniform along its perpendicular direction. The BdG equations are solved in the momentum space so that certain local inhomogeneous feature is neglected. This is also different from our model and actually the gap fluctuation is also an important part of the present work. Some interesting signatures for the phase transition are provided based on the present work which may be detected by later experiments.
Ref.~\cite{chqu} mainly discussed the numerical results for the pure one-dimensional system and weakly coupled chains. Actually,  the topological phase near $\mu=0$ region could not appear for the one-dimensional model.
 It appears near $\mu=-2$, which is the bottom of the normal state band. Thus the system is also a one-band system, and a direct transition from the FFLO state to the topological state can hardly occur.

\begin{figure}
\centering
  \includegraphics[width=3.4in]{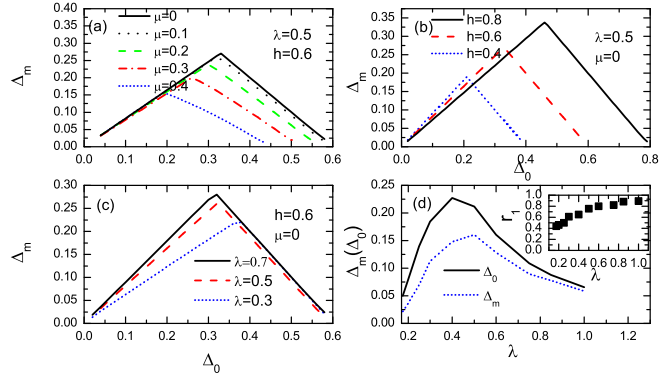}
\caption{The numerical results of the minigap for the $400\times3$ lattice with different parameters. Here the bulk gap $\Delta_0$ is defined as the gap in the site $(200,2)$. (a-c) The minigap as a function of the bulk gap. (d) The minigap and bulk gap as a function of the spin-orbital coupling strength.
Inset of (d): the ratio of the minigap to the bulk gap.}
\end{figure}

It was proposed that the topological order may form before the disappearance of the FFLO state.
Thus a special region, named as "topological FFLO state" might
appear in the phase diagram~\cite{kseo,cqu,zha}. Usually the topological order can be determined strictly through studying the Pfaffian topological invariant.
In Refs.~\cite{kseo,cqu,zha}, the model includes both in-plane Zeeman field and $c$-axis Zeeman field. Such a model would favor an FF state  with $\Delta({\bf i})=\Delta_0 \exp(i{\bf Q}\cdot \bf{R_i})$, in which
the order parameter magnitude is spatially uniform. The phase factor in the order parameter can be gauged out and the BdG hamiltonian can be transformed to the momentum space~\cite{cqu}. This technique can be used to investigate the Pfaffian topological invariant, and a topological FF state has indeed been verified numerically~\cite{kseo,cqu}. In the present work, only $c$-axis Zeeman field is considered.
The quasiparticle bands are degenerate with the momentum ${\bf k}$ and ${\bf -k}$.
For this case, usually LO state with $\Delta({\bf i})=\Delta_0 \cos ({\bf Q}\cdot \bf{R_i})$ is favored. The translational symmetry is broken and thus it is difficult to study numerically the possible topological order in the LO state. While if the topological phase could persist into the LO state,
the possible "topological-LO" state is quite interesting, which may be worthwhile studying in future.

It is also meaningful to compare our results with previous numerical and analytical results on the topological states for the continuous model.
A main character for the topological state is the existence of
 the zero energy states protected by a minigap.
The magnitude of the minigap is an important parameter that can determine the robustness of the MF state against various fluctuations and disorder/impurity effect. Generally the minigap should depend strongly on the magnitude of the bulk pairing gap.
With a uniform pairing gap, the minigap can  approximately be obtained in the momentum space. Namely, the Hamiltonian [Eq.(4) in the method section] can be rewritten in the momentum space considering periodic boundaries. The quasiparticle bands for the Hamiltonian can be obtained numerically, denoted as $E_i({\bf k})$.
The quasiparticle bands are fully gapped and
the minigap should be the minimum value of $\mid E_i({\bf k}) \mid$. In the present work, we solve Eq.(4) in a fully self-consistent method and obtain the minigaps numerically in the real space. For a very large lattice-size,  the same result is obtained for the two methods. In real space, the gap is size dependent. We define the bulk gap $\Delta_0$ to be the gap at the site $(200,2)$. Numerically as the other parameters ($h$, $\lambda$, $\mu$) are fixed, the bulk gap can be controlled through varying the pairing potential $V$.
We show in Figs.8(a-c) the minigap as a function of the bulk gap. For all of the parameters we considered, the nonzero minigap exists only when the relation $h>\sqrt{\Delta_0^2+\mu^2}$ is satisfied. This numerical conclusion is consistent with previous analytical results~\cite{jdsau,jsau} and numerical ones for a continuous model~\cite{lmao}.
Here
the minigap increases linearly with the bulk gap as the bulk gap is small and then decreases linearly for large bulk gap.
This result is also qualitatively consistent with that of the continuous model~\cite{rwei,jdsau,jsau,lmao}.
The scale coefficient for small bulk gap almost does not depend on the chemical potential [Fig.8(a)]. It decreases slightly as the exchange field increases [Fig.8(b)]. While it depends strongly on the spin-orbital strength $\lambda$, as seen in Fig.8(c). It increases as $\lambda$ increases. The relation between the minigap/the bulk gap) and the spin-orbital strength is presented in Fig.8(d). The ratio of the minigap to the bulk gap with a fixed pairing potential $V=2.4$ is plotted in the inset of Fig.8(d). As is seen, the ratio increases monotonously with $\lambda$ and saturates to about 1.0 at a strong limit of spin-orbital coupling. While the magnitudes of the bulk gap and minigap are non-monotonous as the spin-orbital strength increases. The minigap reaches the maximum for about $\lambda=0.4$.

Finally, we would like to discuss how  the present work is relevant to a real system.
Generally, our model might be simulated by cold atoms.
Recently, the spin-orbital coupling has been realized in one-dimensional systems with equal Rashba and Dresselhaus strengths~\cite{lin,wang,cheu}.
While the pure Rashba-type spin-orbital coupling is used in the present work.
Although the experimental realization of the Rashba-type spin-orbital coupling is still awaited,
 there have been several theoretical proposals to generate it
in cold atoms very recently~\cite{cam,jdsa,xuzf}, which seem to be quite promising.  The $s$-wave pairing superfluid state has also been realized in cold atoms~\cite{bou}. We also wish to point out that there always exist the external trap in cold atom systems, which was addressed in Ref.~\cite{TZhou} but is not considered in the present work. The effect of the trap is also interesting and may deserve a further study. While, in principle, our main results are robust again small fluctuations, and we expect that the existence of external trap does not change our main conclusions. At last, we emphasize that we here mainly focus on the phase transition in the lattice system. The numerical results are rather different from those obtained with a continuous model. The existence of the lattice seems to play a rather important role to reach our main conclusion.

In summary, we have studied the physical properties and competing superconducting phase based on a lattice model that includes the $s$-wave pairing, spin-orbital coupling, and the Zeeman field term. The phase transition from the FFLO state to the TSC state induced by the spin orbital coupling has been revealed and discussed. The difference between our lattice model and a previous continuous model are discussed in detail. It has been proposed that the transition from the FFLO state to topological state is easier to occur for the lattice model.
We have also explored the properties of the critical point, the minigap, and the excitations of MFs in the TSC state, providing a helpful insight in profound understanding of topological superconductivity and potential applications. The LDOS spectra have been calculated, serving as a link between our theoretical analysis and experimental observations.

\vspace{3ex}

\noindent
\textbf{Methods}

\noindent
The Hamiltonian can be diagonalized by solving the Bogoliubov-de
Gennes (BdG) equations as
\begin{equation}
\left(
\begin{array}{cc}
 H_t({\bf r}) & \Delta({{\bf r}})\sigma_3  \\
 \Delta^{*}({{\bf r}})\sigma_3 & -\sigma_2 H^{*}_{t}({\bf r}) {\sigma_2}
\end{array}
\right)  \begin{array}{c} \Psi^{n}({\bf r})
\end{array}
 =E_n \begin{array}{c}
\Psi^{n}({\bf r})
\end{array},
\end{equation}
where $\Psi^{n}({\bf r})$ denotes the order column vector with $\Psi^{n}({\bf
r})=(u^{n}_{{\bf r}\uparrow},u^{n}_{{\bf r}\downarrow},v^{n}_{{\bf
r}\downarrow},v^{n}_{{\bf r}\uparrow})^{\mathrm{T}}$, and
the order parameter $\Delta({\bf r})$ is determined
self-consistently
\begin{eqnarray}
\Delta({\bf r})=\frac{V}{2}\sum_n u^{n}_{{\bf
r}\uparrow}v^{n*}_{{\bf r}\downarrow}\tanh (\frac{E_n}{2K_B T})
\end{eqnarray}
with $V$ the pairing strength.

Considering the order
parameter to be uniform along the $y$-direction,
the two-dimensional Hamiltonian can be reduced to a quasi-one dimensional model,
expressed as,
\begin{eqnarray}
H=&\sum_{x,k_y}[\psi^ \dagger_{k_y}(x)(\varepsilon(k_y)\sigma_0+
2\lambda\sin k_y\sigma_2-h\sigma_3) \psi_{k_y}(x)\nonumber\\&-\psi^
\dagger_{k_y}(x)T
\psi_{k_y}(x+1)-h.c.]\nonumber\\&+\sum_{x,k_y}\Delta(x)
\psi^\dagger_{k_y}(x) i\sigma_2 \psi^\dagger_{k_y}(x)+h.c.,
\end{eqnarray}
where $\varepsilon(k_y)=-2 \cos k_y-\mu$ and
$T={{\sigma_0}-i\lambda{\sigma_1}}$.

The BdG equation and self-consistently determined order parameter
are written as,
\begin{equation}
\left(
\begin{array}{cc}
 H_t(k_y,x) & \Delta(x)\sigma_3  \\
 \Delta^{*}(x)\sigma_3 & -\sigma_2 H^{*}_{t}(k_y,x) {\sigma_2}
\end{array}
\right)
\begin{array}{c} \Psi^{n}(k_y,x)
\end{array}
 =E_n \begin{array}{c}
\Psi^{n}(k_y,x)
\end{array}.
\end{equation}

\begin{eqnarray}
\Delta(x)=\frac{V}{2N_y}\sum_{k_y,n} u^{n}(k_y,x)v^{n*}(k_y,x)\tanh
(\frac{E_n}{2K_B T}).
\end{eqnarray}

The local density of states (LDOS) can be calculated as
\begin{equation}
\rho_{\bf i}(\omega)=\sum_n [\mid u^{n}_{{\bf i}\uparrow} \mid^2
\delta(E_n-\omega)+\mid v^{n}_{{\bf i}\downarrow} \mid^2
\delta(E_n+\omega)].
\end{equation}
The delta function $\delta(E)$ is taken as
$\delta=x/[\pi(E^2+x^2)$], with the quasiparticle damping
$x=0.01$.

\noindent
\textbf{Acknowledgments}

\noindent
This work was supported
by the NSFC (Grant No. 11374005 and No. 11204138), the NCET (Grant No. NCET-12-0626), NSF of Jiangsu Province of China (Grant No. BK2012450), the GRF (Grant No. HKU7058/11P\&HKU7045/13P) and the CRF (Grant No. HKU8/11G) of Hong Kong.

\noindent
\textbf{Author contributions}

\noindent
T.Z. performed the numerical calculations and analyzed the data. Y.G. joined in the data analyses. Z.D.W. supervise the whole work. All of the authors
contributed to data interpretation and the writing of the manuscript.

\noindent
\textbf{Additional information}

\noindent
Competing financial interests: The authors declare no competing financial interests.

\noindent
\textbf{Figure Legends}

Fig.1: The topological feature for a two dimensional system. (a) and (b) are the numerical results for the energy spectra with different spin orbital coupling strengths, respectively. The open and periodic boundary condition along the $x$- and $y$- directions are considered.
(c) and (d) are the corresponding order parameters. The other parameters are $\mu=0$ and $h=0.6$.

Fig.2: The low energy spectrum for the quasi-one-dimensional system with the open boundary condition along the $x$-direction. $n$ is the index of eigenstates. Two zero energy eigenvalues can be seen clearly, protected by a minigap $\Delta_m$. The parameters are $\lambda=0.5$, $\mu=0$, and $h=0.6$.

Fig.3: The numerical results of the order parameter $\Delta$ in different phases for the $400\times 3$ system.
(a) The order parameter in the FFLO state. (b) The order parameter at the critical point. (c-d) The order parameter in the topological state.
The parameters are $h=0.6$ and $\mu=0$.

Fig.4: The spatial distributions of the two MF states with different spin-orbital coupling strengths.
(a) The MF states at the critical point where the FFLO modulation disappears. (b) The oscillated MF states in the topological state.
(c) The spatial distribution of the MF states when entering deeply into the topological state. The parameters are $h=0.6$ and $\mu=0$.

Fig.5: The numerical results of the LDOS. (a) The LDOS as a function of the energy. The zero energy states are indicated by the arrows. (b-d) The LDOS as a function of the spatial position with different spin-orbital coupling strengths. The parameters are $h=0.4$ and $\mu=0$.

Fig.6: The quasiparticle band dispersion [Eq.(3)] along the cuts $k_y=0$ and $k_y=2\pi/3$, respectively. The parameters are $\lambda=0.5$, $h=0.6$, and $\mu=0$.

Fig.7: The topological state for the chemical potential near $-4$ and $0$. (a-c): The energy spectra for the $100\times100$ two dimensional lattice with different chemical potentials. (d-f): The quasiparticle energies for the $400\times3$ lattice. The open and periodic boundaries are considered along the $x$ and $y$ directions, respectively. The parameters are set as $h=0.6$, $\Delta \equiv 0.3$, and $\lambda=0.5$.

Fig.8: The numerical results of the minigap for the $400\times3$ lattice with different parameters. Here the bulk gap $\Delta_0$ is defined as the gap in the site $(200,2)$. (a-c) The minigap as a function of the bulk gap. (d) The minigap and bulk gap as a function of the spin-orbital coupling strength.
Inset of (d): the ratio of the minigap to the bulk gap.


\begin{thebibliography}{99}
\bibitem{qi} Qi, X. L. \& Zhang, S. C. Topological insulators and superconductors. {\it Rev. Mod. Phys.} {\bf 83}, 1057-1110 (2011).
\bibitem{roy} Roy R. Topological superfluids with time reversal symmetry. arXiv:0803.2868 (unpublished).
\bibitem{sch} Schnyder, A. P., Ryu, S., Furusaki, A. \& Ludwig, Andreas W. W. Classification of topological insulators and superconductors in three spatial dimensions. {\it Phys. Rev. B} {\bf 78}, 195125 (2008).
\bibitem{qia} Qi, X. L., Hughes, T. L., Raghu, S. \& Zhang, S. C. Time-Reversal-Invariant Topological Superconductors and Superfluids in Two and Three Dimensions. {\it Phys. Rev. Lett.} {\bf 102}, 187001 (2009).
\bibitem{kit} Kitaev, A. Periodic table for topological insulators and superconductors. {\it AIP Conf. Proc.} {\bf 1134}, 22-26 (2009).
\bibitem{hor} Hor, Y. S. {\it et al.} 
Superconductivity in Cu$_x$Bi$_2$Se$_3$ and its Implications for Pairing in the Undoped Topological Insulator.
{\it Phys. Rev. Lett.} {\bf 104}, 057001 (2010).
\bibitem{wra} Wray, L. A. {\it et al.} 
Observation of topological order in a superconducting doped topological insulator.
{\it Nat. Phys.} {\bf 6}, 855-859 (2010).
\bibitem{andr} Wray, L. A. {\it et al.} 
    Spin-orbital ground states of superconducting doped topological insulators: A Majorana platform.
    {\it Phys. Rev. B} {\bf 83}, 224516 (2011).
\bibitem{sasa} Sasaki, S. {\it et al.} 
Topological Superconductivity in Cu$_x$Bi$_2$Se$_3$.
{\it Phys. Rev. Lett.} {\bf 107}, 217001 (2011).
\bibitem{lfu} Fu, L \& Kane, C. L. Superconducting Proximity Effect and Majorana Fermions at the Surface of a Topological Insulator. {\it Phys. Rev. Lett.} {\bf 100}, 096407 (2008).

\bibitem{sato} Sato, M., Takahashi, Y. \& Fujimoto, S. Non-Abelian Topological Order in s-Wave Superfluids of Ultracold Fermionic Atoms. {\it Phys. Rev. Lett.} {\bf 103}, 020401 (2009).
\bibitem{zhcw} Zhang, C., Tewari, S., Lutchyn, R. M. \& Sarma, S. D. $p_x+ip_y$ Superfluid from s-Wave Interactions of Fermionic Cold Atoms. {\it Phys. Rev. Lett.} {\bf 101}, 160401 (2008).
\bibitem{jliu} Liu, J., Han, Q., Shao, L. B. \& Wang, Z. D. Exact Solutions for a Type of Electron Pairing Model with Spin-Orbit Interactions and Zeeman Coupling. {\it Phys. Rev. Lett.} {\bf 107}, 026405 (2011).
\bibitem{rwei} Wei, R. \& Mueller, E. J. Majorana fermions in one-dimensional spin-orbit-coupled Fermi gases. {\it Phys. Rev. A} {\bf 86}, 063604 (2012).
\bibitem{bou}  Bourdel, T. {\it et al.} Experimental Study of the BEC-BCS Crossover Region in Lithium 6.
    {\it Phys. Rev. Lett.} {\bf 93}, 050401 (2004).

    \bibitem{chin} Chin, J. K. {\it et al.}
    Evidence for superfluidity of ultracold fermions in an optical lattice.
    {\it Nature} {\bf 443}, 961-964 (2006).
\bibitem{lin} Lin, Y. J., Jimenez-Garcia, K. \& Spielman, I. B.
Spin-orbit-coupled Bose-Einstein condensates. {\it Nature} {\bf 471}, 83-86 (2011).
\bibitem{wang} Wang, P. {\it et al.} 
Spin-Orbit Coupled Degenerate Fermi Gases.
{\it Phys. Rev. Lett.} {\bf 109}, 095301 (2012).
\bibitem{cheu} Cheuk, L. W. {\it et al.} 
 Spin-Injection Spectroscopy of a Spin-Orbit Coupled Fermi Gas.
{\it Phys. Rev. Lett.} {\bf 109},
095302 (2012).
\bibitem{lutc} Lutchyn, R. M., Sau, J. D. \& Sarma, S. D. Majorana Fermions and a Topological Phase Transition
in Semiconductor-Superconductor Heterostructures. {\it Phys. Rev. Lett.} {\bf 105}, 077001 (2010).
\bibitem{oreg} Oreg, Y., Refael, G. \& Oppen, F. V. Helical Liquids and Majorana Bound States in Quantum Wires. {\it Phys. Rev. Lett.} {\bf 105}, 177002 (2010).
\bibitem{wil} Williams, J. R. {\it et al.} 
    Unconventional Josephson Effect in Hybrid Superconductor-Topological Insulator Devices.
    {\it Phys. Rev. Lett.} {\bf 109}, 056803 (2012).
\bibitem{rok} Rokhinson, L. P., Liu, X. \& Furdyna., J. K., Observation of the fractional ac Josephson effect: the signature of Majorana particles. {\it Nat. Phys.} {\bf 8}, 795-799 (2012).
\bibitem{den} Deng, M. T. {\it et al.} 
Anomalous Zero-Bias Conductance Peak in a Nb-InSb Nanowire-Nb Hybrid Device.
{\it Nano Lett.} {\bf 12}, 6414-6419 (2012).
\bibitem{das} Das, A. {\it et al.} 
Zero-bias peaks and splitting in an Al-InAs nanowire topological superconductor as a signature of Majorana fermions.
{\it Nat. Phys.} {\bf 8}, 887-895 (2012).
 \bibitem{mou} Mourik, V. {\it et al.} 
 Signatures of Majorana Fermions in Hybrid Superconductor-Semiconductor Nanowire Devices.
 {\it Science} {\bf 336}, 1003-1007 (2012).
\bibitem{ful} Fulde, P. \& Ferrell, R. A. Superconductivity in a Strong Spin-Exchange Field. {\it Phys. Rev.} {\bf 135}, A550-A563 (1964).
\bibitem{lar} Larkin A. I. \& Ovchinnikov, Y. N. Inhomogeneous State of Superconductors. {\it Sov. Phys. JETP} {\bf 20}, 762-769 (1965).
\bibitem{asl} Aslamazov, L. G. Influence of Impurities on the Existence of an Inhomogeneous State in a Ferromagnetic Superconductor. {\it Sov.
Phys. JETP} {\bf 28}, 773-775 (1969).
\bibitem{grue} Gruenberg. L. W. \& Gunther L., Fulde-Ferrell Effect in Type-II Superconductors. {\it Phys. Rev. Lett.} {\bf 16}, 996-998 (1966).

\bibitem{rado} Radovan, H. A. {\it et al.} 
Magnetic enhancement of superconductivity from electron spin domains.
{\it Nature} {\bf 425}, 51-55 (2003).
\bibitem{bia} Bianchi, A., Movshovich, R., Capan, C., Pagliuso, P. G. \& Sarrao, J. L.
Possible Fulde-Ferrell-Larkin-Ovchinnikov Superconducting State in CeCoIn$_5$.
{\it Phys. Rev. Lett.} {\bf 91}, 187004 (2003).

\bibitem{kum} Kumagai, k. {\it et al.} 
Fulde-Ferrell-Larkin-Ovchinnikov State in a Perpendicular Field of Quasi-Two-Dimensional CeCoIn$_5$.
{\it Phys. Rev. Lett.} {\bf 97}, 227002 (2006).

\bibitem{tan} Tanatar, M. A., Ishiguro, T., Tanaka, H. \&
Kobayashi, H. Magnetic field-temperature phase diagram of the quasi-two-dimensional organic superconductor ¦Ë-(BETS)$_2$GaCl$_4$ studied via thermal conductivity.
{\it Phys. Rev. B} {\bf 66}, 134503 (2002).


\bibitem{uji} Uji, S. {\it et al.} Magnetic-field-induced superconductivity in a two-dimensional organic conductor. {\it Nature} {\bf
410}, 908-910 (2001).
\bibitem{bal} Balicas, L. {\it et al.} Superconductivity in an Organic Insulator at Very High Magnetic Fields. {\it Phys. Rev. Lett.} {\bf 87}, 067002 (2001).
\bibitem{shi} Shimahara, H. Fulde-Ferrell-Larkin-Ovchinnikov State in a Quasi-Two-Dimensional Organic Superconductor. {\it J. Phys. Soc. Jpn.} {\bf 66}, 541-544 (1997).
\bibitem{man} Manalo, S. \& Klein, U. Has the Fulde-Ferrell-Larkin-Ovchinnikov state been observed in the organic superconductor ¦Ê-(BEDT-TTF)$_2$Cu(NCS)$_2$? {\it J. Phys.: Condens. Matter} {\bf 12},
L471-L476 (2000).
\bibitem{liao} Liao, Y. {\it et al.} Spin-imbalance in a one-dimensional Fermi gas. {\it Nature} {\bf 467}, 567-569 (2010).
\bibitem{han-wang} Han, Q., Liu, J., Zhang, D. \& Wang, Z. D. Unconventional Fulde-Ferrel-Larkin-Ovchinnikov states in spin-orbit coupled condensates: exact results. arXiv: 1104.0614 (unpublished).
    \bibitem{kseo} Seo, K., Zhang, C., \& Tewari S. Topological uniform superfluid and Fulde-Ferrell-Larkin-Ovchinnikov phases in three-dimensional
to one-dimensional crossover of spin-orbit-coupled Fermi gases. {\it Phys. Rev. A} {\bf 88}, 063601 (2013).
    \bibitem{chqu} Qu, C., Gong, M., \& Zhang, C. FFLO or Majorana superfluids: The fate of fermionic cold atoms in spin-orbit coupled
optical lattices. arXiv:1304.3926 (unpublished).
    \bibitem{yxu} Xu, Y., Qu, C., Gong, M, \& Zhang, C. Competing superfluid orders in spin-orbit coupled fermionic cold atom optical lattices. arXiv:1305.2152 (unpublished).
\bibitem{cqu} Qu, C. {\it et al.} Topological Superfluids with Finite Momentum Pairing and Majorana Fermions. {\it Nat. Commun.} {\bf 4}, 2710 (2013).

\bibitem{zha} Zhang W. \& Yi, W. Topological Fulde-Ferrell-Larkin-Ovchinnikov states in spin¨Corbit-coupled Fermi gases. {\it Nat. Commun.} {\bf 4}, 2711 (2013).
\bibitem{fwu} Wu, F., Guo, G. C., Zhang, W. \& Yi, W. Unconventional Fulde-Ferrell-Larkin-Ovchinnikov pairing states in a Fermi gas with spin-orbit coupling. {\it Phys. Rev. A} {\bf 88}, 043614 (2013).
\bibitem{lmao} Mao L. \& Zhang, C. Robustness of Majorana modes and minigaps in a spin-orbit-coupled semiconductor-superconductor heterostructure. {\it Phys. Rev. B} {\bf 82}, 174506 (2010).
\bibitem{note} We have also checked numerically the size effect and this result does not change for a larger system size $(1000\times 3)$.


\bibitem{jdsau} Sau, J. D., Lutchyn, R. M., Tewari, S. \& Sarma, S. D. Generic New Platform for Topological Quantum Computation Using Semiconductor Heterostructures. {\it Phys. Rev. Lett.} {\bf 104}, 040502 (2010).
   \bibitem{jsau} Sau, J. D., Tewari, S., Lutchyn, R. M., Stanescu, T. D. \& Sarma, S. D. Non-Abelian quantum order in spin-orbit-coupled semiconductors: Search for topological Majorana particles in solid-state systems. {\it Phys. Rev. B} {\bf 82}, 214509 (2010).
       \bibitem{alic} Alicea, J., Majorana fermions in a tunable semiconductor device. {\it Phys. Rev. B} {\bf 81}, 125318 (2010).

       \bibitem{kyang} Yang, K. Inhomogeneous superconducting state in quasi-one-dimensional systems. {\it Phys. Rev. B} {\bf 63}, R140511 (2001).
\bibitem{fid} Fidkowski, L., Lutchyn, R. M., Nayak, C. \& Fisher, M. P. A. Majorana zero modes in one-dimensional quantum wires without long-ranged superconducting order. {\it Phys. Rev. B} {\bf 84}, 195436 (2011).
\bibitem{jayd} Sau, J. D., Halperin, B. I.,  Flensberg, K. \&  Sarma., S. D. Number conserving theory for topologically protected degeneracy in one-dimensional fermions. {\it Phys. Rev. B} {\bf 84}, 144509 (2011).
\bibitem{cam} Campbell, D. L., Juzeliunas, G. \& Spielman, I. B. Realistic Rashba and Dresselhaus spin-orbit coupling for neutral atoms. {\it Phys. Rev. A} {\bf 84}, 025602 (2011).
\bibitem{jdsa} Sau, J. D. {\it et al.} Chiral Rashba spin textures in ultracold Fermi gases. {\it Phys. Rev. B} {\bf 83}, R140510
(2011).
\bibitem{xuzf} Xu, Z. F. \& You, L. Dynamical generation of arbitrary spin-orbit couplings for neutral atoms. {\it Phys. Rev. A} {\bf 85}, 043605 (2012).
\bibitem{TZhou} Zhou, T. \& Wang, Z. D. Revealing Majorana fermion states in a superfluid of cold atoms subject to a harmonic potential. {\it Phys. Rev. B} {\bf 88}, 155114 (2013).

\end{thebibliography}
\end{document}